# Tracing Technique for Blaster Attack

Siti Rahayu S., Robiah Y., Shahrin S., Faizal M. A., Mohd Zaki M, Irda R.

Faculty of Information Technology and Communication

Univeristi Teknikal Malaysia Melaka,

Durian Tunggal, Melaka,

Malaysia

sitirahayu@utem.edu.my, robiah@utem.edu.my, shahrinsahib@utem.edu.my,

faizalabdollah@utem.edu.my,zaki.masud@utem.edu.my, irda@utem.edu.my

*Abstract -* **Blaster worm of 2003 is still persistent, the infection appears to have successfully transitioned to new hosts as the original systems are cleaned or shut off, suggesting that the Blaster worm, and other similar worms, will remain significant Internet threats for many years after their initial release. This paper is to propose technique on tracing the Blaster attack from various logs in different OSI layers based on fingerprint of Blaster attack on victim logs, attacker logs and IDS alert log. The researchers intended to do a preliminary investigation upon this particular attack so that it can be used for further research in alert correlation and computer forensic investigation.**

*Keyword; Tracing technique, Blaster attack, fingerprint, log*

## I. INTRODUCTION

The Blaster worm of 2003 infected at least 100,000 Microsoft Windows systems and cost millions in damage. In spite of cleanup efforts, an antiworm, and a removal tool from Microsoft, the worm persists [1]. According to [2], research on Blaster attack is significant due to the multitude of malware such as Blaster worm has itself evolved into a complex environment and has potential for reinfection by either itself or another worm, to occur using the same exploit.

Recent tools targeted at eradicating it appear to have had little effect on the global population. In the persistent population analysis, the infection appears to have successfully transitioned to new hosts as the original systems are cleaned or shut off, suggesting that the Blaster worm, and other similar worms, will remain significant Internet threats for many years after their initial release and its suggested that the Blaster worm is not going away anytime soon. Therefore, the objective of this paper is to propose technique on tracing the Blaster attack from various logs in different OSI layers. The researchers intended to do a preliminary investigation upon this particular attack so that it can be used for further research in alert correlation and computer forensic investigation.

## II. RELATED WORK

W32.Blaster.Worm is a worm that exploits the DCOM RPC vulnerability (described in Microsoft Security Bulletin MS03-026) using TCP port 135. If a connection attempt to TCP port 135 is successful, the worm sends an RPC bind command and an RPC request command containing the buffer overflow and exploit code. The exploit opens a backdoor on TCP port 4444, which waits for further commands. The infecting system then issues a command to the newly infected system to transfer the worm binary using Trivial File Transfer Protocol (TFTP) on UDP port 69 from the infecting system and execute it.

The worm targets only Windows 2000 and Windows XP machines. While Windows NT and Windows 2003 Server machines are vulnerable to the aforementioned exploit (if not properly patched), the worm is not coded to replicate to those systems. This worm attempts to download the msblast.exe file to the %WinDir%\system32 directory and then execute it.

The Blaster worm's impact was not limited to a short period in August 2003. According to [3], a published survey of 19 research universities showed that each spent an average of US$299,579 during a five-week period to recover from the Blaster worm and its variants. The cost of this cleanup effort has helped solidify a growing view of worms not as acts of Internet vandalism but as serious crimes. Although the original Blaster.A author was never caught, authors of several other variants have been apprehended.

There are various research techniques done by others researcher in detecting attack. It can either use signature-based, anomaly-based or specification-based. The signature-based as described by [4] will maintain the database of known intrusion technique and detects intrusion by comparing behaviour against the database whereas the anomaly-based detection techniques will analyses user behaviour and the statistics of a process in normal situation, and it checks whether the system is being used in a different manner. [5] has described that this technique can overcome misuse detection problem by focusing on normal system behaviour rather than attack behaviour. The specification-based detection according to [6] will rely on program specifications that describe the intended behaviour of security-critical programs. The research trend for detecting attack has move towards combination or hybrid of either signature-based with anomaly-based done by [7], [8] and [5] or specification-based with anomaly-based done by [9].

For the purpose of this preliminary experiment, the researchers have selected only signature-based detection technique and in future, intend to combine it with anomaly-based detection technique for further improvement of tracing attack.

System log files contain valuable evidence pertaining to computer attacks. However, the log files are often massive, and much of the information they contain is not relevant to the network administrator. Furthermore, the files almost always have a flat structure, which limits the ability to query them. Thus, it is extremely difficult and time consuming to extract and analyse the trace of attacks from log files [10]. This paper will select the most valuable attributes from a log file that is relevance to the attack being traced. Our research is preliminary experiment of tracing the Blaster.B attack in diverse log resources to provide more complete coverage of the attack space [11].

According to [12], the network attack analysis process involves three main procedures: initial response, media imaging duplication, and imaged media analysis. Our proposed approach focuses on the procedure of media imaging duplication and imaged media analysis. This paper describes how procedure can be applied to the numerous logs, which can derive the top facts in each of the diverse connections and locate malicious events spread across the network.

## III. EXPERIMENT APPROACH

Our proposed approach in this preliminary experiment used four methods: Network Environment Setup, Attack Activation, Log Collection and Log Analysis and its depicted in Figure 1. The details of the method are discussed in the following sub-section.

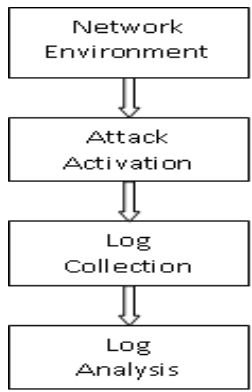

Figure 1: Method use in the preliminary experiment

### A. Network Environment Setup

The network setup for this experiment will refer to the network simulation setup [13] done by the MIT Lincoln Lab and it has been slightly modified using only *Centos* and *Windows XP* compared to MIT Lincoln Lab which using *Linux, Windows NT, SunOS, Solaris, MacOS* and *Win98* to suit our experiment's environment. The network design is as shown below in Figure 2.

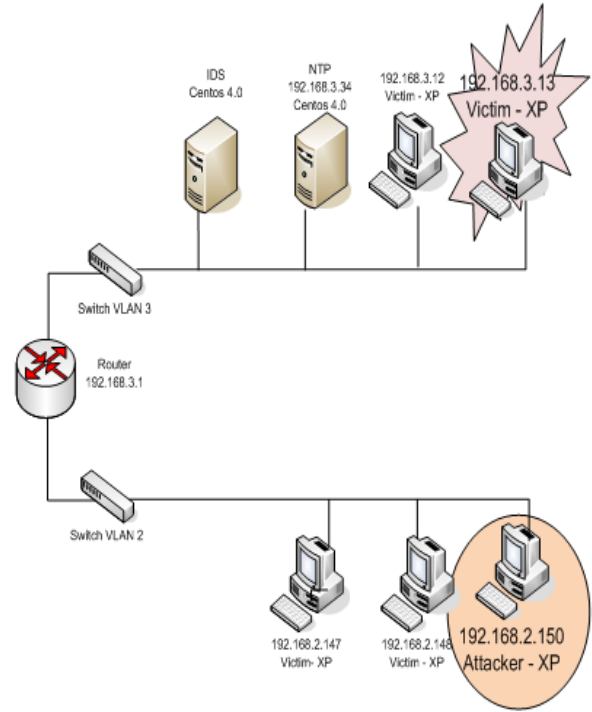

Figure 2: Preliminary Network Design for Blaster Attack Simulation

This network design consists of two switches configured to Vlan 3 (192.168.3.0) and Vlan 2 (192.168.2.0), one router, two servers for Intrusion Detection System (IDS) and Network Time Protocol (NTP) run on *Centos 4.0*, two victims run on *Windows XP* on each Vlan and one attacker run on Vlan 2. The log files that expected to be analysed are four types of log files (*personal firewall log, security log, system log* and *application log*) that shall be generated by host level device and one log files by network level device (*alert log* by IDS). Ethereal 0.10.7 [6] were installed in each host to verify the traffic between particular host and other device and *tcpdump* script is activated in IDS to capture the traffic for the whole traffic within Vlan 2 and Vlan 3.

### B. Attack Activation

Event viewer and time synchronisation using NTP server is configured before attack is launched. Then Blaster variant is installed and activated on the attacker machine. This experiment runs for 30 minutes. Once the victim machine is successfully infected by the Blaster, the experiment is terminated.

### C. Log Collection

Log is collected at two different OSI layers which are application layer and network layer. Each victim and attacker machine will generated *personal firewall log, security log, application log, system log* and *ethereal log*. The IDS machine will generate *alert* and *tcpdump log*. *Ethereal* and *tcpdump* files are used to verify the simulation attack and compare it with the others log files. For the purpose of this paper, both verification logs are not discussed due to limited page. The summary of the various log files generated is as shown in Table I.

TABLE I. Various log files generated from two different OSI layers

| OSI layers | Log Filename | Description | Log Location |
|---|---|---|---|
| Application | pfirewall.log | Personal firewall log | victim and attacker |
| | security.evt | Security log in event viewer | victim and attacker |
| | application.evt | Application log in event viewer | victim and attacker |
| | system.evt | System log in event viewer | victim and attacker |
| Network | alert.log | Alert log from IDS | IDS server |
| | tcpdump | tcpdump log to capture overall network traffic | IDS server |
| | ethereal | Network traffic capture at specific host only. | victim and attacker |

### C. Log Analysis

In this network attack analysis process the researchers has implement the media imaging duplication using IDS and imaged media analysis by analysing logs generated in Table 1. The objective of the log analysis is to identify the Blaster attack by observing the specific characteristics of the Blaster attack which exploits the DCOM RPC vulnerability using TCP port 135. This worm attempts to download the msblast.exe file to the *%WinDir%\system32* directory and then execute it. The exploit opens a backdoor on TCP port 4444, which waits for further commands. In this analysis, the researchers have selected the valuable attributes that is significance to the attack being traced as shown in Table II.

TABLE II. Selected Log Attribute

| Log filenames | Selected Log Attribute | Variable |
|---|---|---|
| pfirewall.log | • Source IP address<br>• Destination IP Address<br>• Destination port<br>• Source port<br>• Action<br>• Date<br>• Time | • SrcIP<br>• DstIP<br><br>• Dstport<br>• Srcport<br>• Act<br>• D<br>• T |
| security.evt<br>application.evt<br>system.evt | • Date<br>• Time<br>• Category | • D<br>• T<br>• Cat |
| alert.log | • Date<br>• Time<br>• Source IP address<br>• Destination IP Address<br>• Category | • D<br>• T<br>• SrcIP<br>• DstIP<br><br>• Cat |

## IV. PROPOSED TRACING TECHNIQUE

In order to identify the attacker, the researchers have proposed a tracing technique as depicted in Figure 3, consists of three elements: victim, attacker and IDS. The algorithm used in each element will be elaborated in the next sub-section.

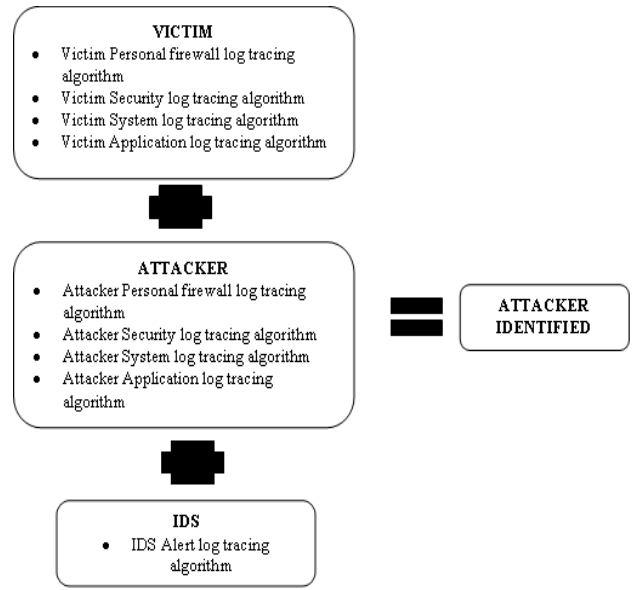

Figure 3: Proposed Tracing Technique

### A. Tracing Algorithm for Victim logs

In our tracing procedure, the tracing activity will be primarily done at victim site by examining the Blaster fingerprint for victim logs as shown in Figure 4. These Blaster fingerprint is derived from several studies done by [14], [15], [16].

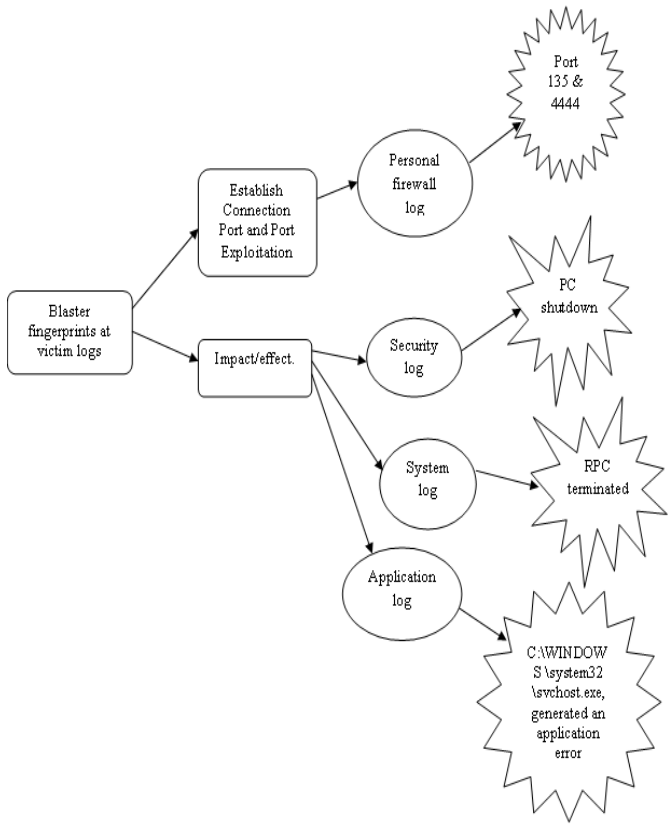

Figure 4: Fingerprint of Blaster attack in each selected victim logs

In this analysis, the researchers have specified 192.168.3.13 as one of the victim and 192.168.2.150 as attacker (refer to Figure 2). The tracing tasks are initially started at the victim *personal firewall log* followed by *security log, system log* and *application log*. The data can be further analysed by referring to Blaster fingerprint for attacker logs by examine the attacker *personal firewall* and *security log*. Figure 6, 9 and 12 is the relevant information that has been extracted from selected logs.

Figure 5 shows the tracing algorithm for each selected victim logs based on Blaster attack fingerprint as in Figure 4.

The aim of these tracing tasks is to examine the trace left by the Blaster in the selected log. The trace is based on the Blaster attack fingerprint which primarily done at *personal firewall log*. In these tracing tasks, the researchers have manipulated the attributes selected in Table II. The searching start with the victim IP address is 192.168.3.13, and the action is OPEN-BOUNDED which show the attacker is trying to open the connection. The protocol used is TCP and the destination port is 135 which show that Blaster attack attempt to establish connection.

```
Where,
    x = Victim Host
    y = Attacker Host
```

**Victim Personal firewall log tracing algorithm**
```
Input Action, Protocol, Destination Port
If (Action = Open-Inbound) and (Protocol = TCP)
and (Destination Port = 135)
   Date = D_FW1 ^x
   Time = T_FW1 ^x
   Source IP = SrcIP^x
   Destination IP = DestIP^x
    Source Port = SrcPort_a^x
   Print Source IP, Date, Time, Source Port,
      Destination IP, Action, Protocol, Destination
      Port
   If (Action = Open) and (Protocol = TCP) and
      (Destination Port = 4444) and (Date = D_FW ^x)
      and (Time >= T_FW1 ^x) and
      (Source IP = SrcIP^x) and (Destination IP =
      DestIP^x)
           Time = T_FW2 ^x
           Source Port = SrcPort_e^x
           Print Source IP, Date, Time, Source
           Port, Destination IP, Action, Protocol,
           Destination Port
   End
End
```

**Victim Security log tracing algorithm**
```
Input Date (D_FW ^x)
Input Time (T_FW2 ^x)
Input AuditCategory
If (Date = D_FW^x) and (Time >= T_FW2^x) and
   (AuditCategory = '\system32\svchost.exe,
   generated an application error')
   Time = T_App1 ^x
   Date = D_App1 ^x
   Print Time, Date, AuditCategory
End
```

**Victim System log tracing algorithm**
```
Input Date (D_App1 ^x)
Input Time (T_App1 ^x)
Input AuditCategory
If (Date = D_App1 ^x) and (Time >= T_App1 ^x) and
   (AuditCategory = 'The Remote Procedure Call
   (RPC) service terminated unexpectedly')
   Time = Time_Sys ^x
   Date = Date_Sys ^x
   Print Time, Date, AuditCategory
End
```

**Victim Application log tracing algorithm**
```
Input Date (D_Sys ^x)
Input Time (T_Sys ^x)
Input AuditCategory
If (Date = D_Sys ^x) and (Time >= T_Sys ^x) and
   (AuditCategory = 'Windows is shutting down')
   Time = Time_Sec ^x
   Date = Date_Sec ^x
   Print Time, Date, AuditCategory
End
```

Figure 5: Tracing algorithm for Victim logs

**Victim Personal firewall log**

```
2009-05-07 14:13:34 OPEN-INBOUND TCP 192.168.2.150
  192.168.3.13 3284 135 - - - - - - -
2009-05-07 14:14:01 DROP TCP 192.168.2.150
  192.168.3.13 3297 4444 48 S 862402054 0 64240 -
  - -
```

**Victim Security log**

```
5/7/2009      2:20:03 PM     Security
    Success Audit System Event 513 NT
    AUTHORITY\SYSTEM    AYU    Windows is shutting
    down. All logon sessions will be terminated by
    this shutdown.
```

**Victim System log**

```
5/7/2009      2:19:00 PM    Service Control
  Manager Error None 7031 N/A AYU
    The Remote Procedure Call (RPC) service
    terminated unexpectedly. It has done this 1
    time(s). The following corrective action will
    be taken in 60000 milliseconds: Reboot the
    machine.
5/7/2009      2:19:00 PM    USER32 Information
  None 1074 NT AUTHORITY\SYSTEM    AYU
    The process winlogon.exe has initiated the
    restart of AYU for the following reason: No
    title for this reason could be found
  Minor Reason: 0xff
  Shutdown Type: reboot
  Comment: Windows must now restart because the
  Remote Procedure Call (RPC) service terminated
  unexpectedly
```

**Victim Application log**

```
5/7/2009      2:20:01 PM    EventSystem    Error
  (50) 4609 N/A AYU    The COM+
  Event System detected a bad return code during
  its internal processing. HRESULT was 800706BA
  from line 44 of
  d:\nt\com\com1x\src\events\tier1\eventsystemobj
  .cpp. Please contact Microsoft Product Support
  Services to report this error.
5/7/2009      2:19:00 PM    DrWatson
  Information None 4097 N/A AYU
  C:\WINDOWS\system32\svchost.exe, generated an
  application error The error occurred on
  05/07/2009 @ 14:19:00.441 The exception
  generated was c0000005 at address 0018759F
  (<nosymbols>)
5/7/2009      2:14:00 PM    Application Error
  Error (100) 1000 N/A AYU
  Faulting application svchost.exe, version
  5.1.2600.0, faulting module unknown, version
  0.0.0.0, fault address 0x00000000.
5/7/2009      2:20:03 PM    EventLog
  Information None 6006 N/A AYU
  The Event log service was stopped.
```

Figure 6: Extracted data from Victim logs

From these trace, the source IP address (SrcIP$^x$) and source port of potential attacker is known where source IP address is *192.168.2.150*, source port (SrcPort$_a$$^x$) is *3824* and the date and time is *2009-05-07 14:13:34* also known to shows when the attack is happen.

Subsequently, to trace whether the attack was exploited, the log is further search on the same date and time within the range of the Blaster attack attempt to establish connection. The destination IP address (DestIP$^x$) is victim IP address, the source IP address (SrcIP$^x$) is the potential attacker IP address, the action is DROP, protocol used is TCP and destination port is 4444. From this trace, the potential attacker source port is known and it indicates that the Blaster is exploited using port 4444. This attack can be further verified by examining the *personal firewall log* at the machine of the potential attacker.

To support the information obtained in *personal firewall log*, further investigation done in the *security log, system log* and *application log*. The effect of the exploitation can be traced by looking at the message embedded in the application log, system log and security log which shows message "*C:\WINDOWS\system32\svchost.exe, generated an application error*", "*Windows must now restart because the Remote Procedure Call (RPC) service terminated unexpectedly*" and "*Windows is shutting down. All logon sessions will be terminated by this shutdown*" respectively. All of these messages shown the effect of Blaster attack, which it exploits the RPC services. The highlighted data in Figure 6 is extracted by using the tracing algorithm in Figure 5 accordingly.

*B. Tracing Algorithm for Attacker logs*

The tracing algorithm for tracing the attacker logs in Figure 8 is based on Blaster attack fingerprint in Figure 7. The same tracing step in victim logs is used in investigating the attacker logs. The only difference is the action is OPEN and extra information obtained from previous tracing tasks: source port (SrcPort$_a$$^x$), date (D$_{FW}$$^x$) and time (T$_{FW1}$$^x$) is used to verify the existence of communications between attacker and victim machine on port 135.

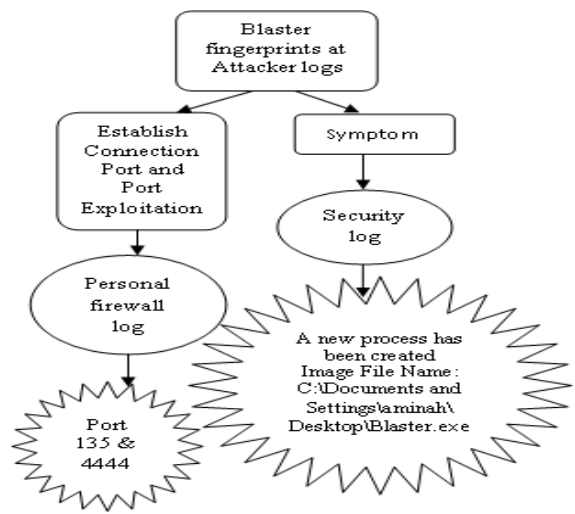

Figure 7: Fingerprint of Blaster attack in each selected attacker log

Then, to verify that there is an exploitation done by attacker to victim machine, the main attributes used in the *personal firewall log* are destination IP address, action is OPEN, protocol is TCP, destination port is 4444, source port (SrcPort$_e^x$), date (D$_{FW}^x$) and time (T$_{FW2}^y$).

To validate the information obtained in the attacker *personal firewall log*, further analysis done in the *security log, system log and application log*. The process created is found in the security log with the message *"A new process has been created and the Image File Name: C:\Documents and Settings\aminah\Desktop\Blaster.exe"*.

```
Where,
       x = Victim Host
       y = Attacker Host
Attacker Personal firewall log tracing algorithm
Input Action, Protocol, Destination Port
Input Date (obtained from tracing victim log,
       D_FW ^x)
Input Time (obtained from firewall victim log,
       T_FW1 ^x)
Input Source IP (obtained from firewall victim
       log, SrcIP^x)
Input Destination IP (obtained from firewall
       victim log, DestIP^x)
Input Source Port to attempt attack (obtained
       from firewall victim log, SrcPort_a^x)
Input Source Port to exploit attack (obtained
       from firewall victim log, SrcPort_e^x)
If (Action = Open) and (Protocol = TCP) and
       (Destination Port = 135) and (Date = D_FW^x)
       and(Time <= T_FW1 ^x) and (Source IP = SrcIP^x)
       and (Destination IP = DestIP^x) and (Source
       Port = SrcPort_a^x)
       Time = T_FW1^y
       Date = D_FW^y
       Print Source IP, Destination IP, Date,
       Time, Source Port, Destination Port,
       Protocol, Action
  If (Action = Open) and (Protocol = TCP) and
       (Destination Port = 4444) and (Date = D_FW^y)
       and (Time >= T_FW1^y) and (Source IP = SrcIP^x)
       and (Destination IP = DestIP^x) and (Source
       Port = SrcPort_e^x)
       Time = T_FW2 ^y
       Print Source IP, Date, Time, Source
       Port,
              Destination IP, Action, Protocol,
              Destination Port
End
End

Attacker Security log tracing algorithm
Input Date (D_FW^y)
Input Time (T_FW2^y)
Input AuditCategory

If (Date = D_FW^y) and (Time >= T_FW2^y) and
   (AuditCategory = 'Windows is shutting down')
       Time = Time_Sec ^y
       Date = Date_Sec ^y
       Print Time, Date, AuditCategory
End
```

Figure 8: Tracing algorithm for Attacker logs

The highlighted data in Figure 9 is extracted by using the tracing algorithm in Figure 8 accordingly.

From the tracing, there is an evidence shows that the attack is launched by this attacker machine (192.168.2.150) at *2009-05-07 14:13:33* which is concurrent with the extracted data in Figure 6. Hence, the attacker can be identified using this tracing algorithm.

**Attacker Personal firewall log**
```
2009-05-07 14:13:33 OPEN TCP 192.168.2.150 192.168.3.12
3283 135 - - - - -
2009-05-07 14:13:33 OPEN TCP 192.168.2.150 192.168.3.13
3284 135 - - - - -
2009-05-07 14:13:33 OPEN TCP 192.168.2.150 192.168.3.14
3285 135 - - - - -
2009-05-07 14:13:33 OPEN TCP 192.168.2.150 192.168.3.15
3286 135 - - - - -
2009-05-07 14:13:35 OPEN TCP 192.168.2.150 192.168.3.12
3296 4444 - - - - -
2009-05-07 14:13:56 OPEN TCP 192.168.2.150 192.168.3.13
3297 4444 - - - - -
2009-05-07 14:14:11 CLOSE TCP 192.168.2.150 192.168.3.12
3283 135 - - - - -
2009-05-07 14:14:11 CLOSE TCP 192.168.2.150 192.168.3.13
3284 135 - - - - -
2009-05-07 14:14:11 CLOSE TCP 192.168.2.150 192.168.3.15
3286 135 - - - - -
2009-05-07 14:15:11 CLOSE TCP 192.168.2.150 192.168.3.12
3296 4444 - - - - -
2009-05-07 14:15:11 CLOSE TCP 192.168.2.150 192.168.3.13
3297 4444 - - - - -
2009-05-07 14:15:11 CLOSE TCP 192.168.2.150 192.168.3.34
3307 135 - - - - -
```

**Attacker Security log**
```
5/7/2009        2:13:08 PM        Security Success Audit
   Detailed Tracing    592    RAHAYU2\aminah
   RAHAYU2  A new process has been created:
   New Process ID:   1640
   Image File Name: C:\Documents and
Settings\aminah\Desktop\Blaster.exe
   Creator Process ID:   844
   User Name:     aminah
   Domain:        RAHAYU2
   Logon ID:              (0x0,0x17744)
```

Figure 9: Extracted data from Attacker logs

*C. Tracing Algorithm for IDS logs*

The Blaster attack fingerprint in Figure 10 is the base for tracing algorithm in IDS alert logs as depicted in Figure 11.

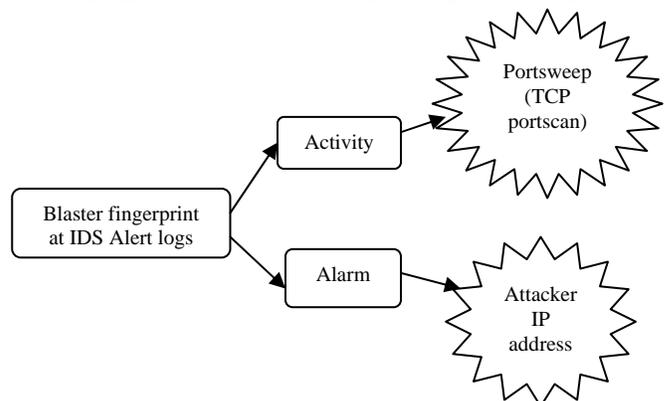

Figure 10: Fingerprint of Blaster attack in IDS log

To confirm that there is an exploitation done by attacker, extra information can be obtained from *IDS alert log*. The main attributes used in the IDS *alert log* are date, time, Source IP Address and destination IP address. If the destination IP address does not exist, the alert has generated false positive alert. However, existence of source IP address is good enough to verify that this source IP address had launched an attack as reported as portsweep activity in IDS alert log shown in Figure 12.

```
Input Date (obtained from victim firewall log,
    D_FW ^x)
Input Start Time (obtained from victim firewall
    log, T_FW1^x)
Input End Time (obtained from victim firewall log,
    T_FW2^x)
Input Source IP (obtained from victim firewall
    log, SrcIP^x)
Input Destination IP (obtained from victim
    firewall log, DestIP^x)
If (Date = D_FW ^x) and (T_FW1^x =<Time>= T_FW2^x) and
    (Source IP = SrcIP^x) and (Destination IP =
    DestIP^x)
    Time = T_IDS
    Print Date, Time, Source IP, Destination IP,
    Alert Message
Else
    If (Date = D_FW ^x) and (T_FW1^x =<Time>= T_FW2^x) and
    (Source IP = SrcIP^x)
    Time = T_IDS
    Print Date, Time, Source IP, Destination IP,
    Alert Message
    End
End
```

Figure 11: IDS tracing algorithm

```
[**] [122:3:0] (portscan) TCP Portsweep [**]
[Priority: 3]
05/07-14:10:56.381141 192.168.2.150 ->
192.168.3.1
PROTO:255 TTL:0 TOS:0x0 ID:14719 IpLen:20
DgmLen:158

[**] [122:3:0] (portscan) TCP Portsweep [**]
[Priority: 3]
05/07-14:11:43.296733 192.168.2.150 ->
192.168.3.34
PROTO:255 TTL:0 TOS:0x0 ID:0 IpLen:20
DgmLen:162 DF
```

Figure 12: Extracted data from IDS alert log

The extracted data depicted from Figure 12, verified that the source IP address (192.168.2.150) is the attacker due to the port scanning alarm generated by the IDS. Thus, all the three tracing algorithm have the capability to identify the attacker.

## V. CONCLUSION AND FUTURE WORKS

In this study, the researchers have reviewed and analysed the Blaster attack from various logs in different OSI layers and researchers' approach focuses on the procedure of media imaging duplication and imaged media analysis. Researchers have selected the most valuable attributes from the log files that are relevance to the attack being traced. From the analysis researcher has propose a technique on tracing the Blaster attack using specific tracing algorithm as in Figure 3 for each log which is based on fingerprint of Blaster attack on victim logs, attackers logs and IDS alert log. This tracing technique is primarily used signature-based technique and later on the researchers intend to merge it with anomaly-based technique to improve the tracing capability. All of these logs are interconnected from one log to another log to provide more complete coverage of the attack space information. Further improvement should be done on generalising the process of detecting the worm attack that will produce attack and trace pattern for alert correlation and computer forensic investigation research.

## VI. REFERENCES

[1]. Bailey, M., Cooke, E., Jahanian, F., Watson, D., & Nazario, J. (2005). The Blaster Worm: Then and Now. IEEE Computer Society

[2]. Crandall, J. R., Ensafi, R., Forrest, S., Ladau, J., & Shebaro, B. (2008). The Ecology of Malware. ACM .

[3]. Foster, A. L. (2004). Colleges Brace for the Next Worm. The Chronicle of Higher Education, 50 (28), A29.

[4]. Okazaki, Y., Sato, I., & Goto, S. (2002). A New Intrusion Detection Method based on Process Profiling. Paper presented at the Symposium on Applications and the Internet (SAINT '02) IEEE.

[5]. Sekar, R., Gupta, A., Frullo, J., Shanbhag, T., Tiware, A., & Yang, H. (2002). Specification-based Anomaly Detection: A New Approach for DetectingNetwork Intrusions. Paper presented at the ACM Computer and Communication Security Conference.

[6]. Ko, C., Ruschitzka, M., & Levitt, K. (1997). Execution monitoring of security critical programs in distributed systems: A Specification-based Approach. Paper presented at the IEEE Symposium on Security and Privacy.

[7]. Bashah, N., Shanmugam, I. B., & Ahmed, A. M. (2005). Hybrid Intelligent Intrusion Detection System. Paper presented at the World Academy of Science, Engineering and Technology, June 2005.

[8]. Garcia-Teodoro, P., E.Diaz-Verdejo, J., Marcia-Fernandez, G., & Sanchez-Casad, L. (2007). Network-based Hybrid Intrusion Detection Honeysystems as Active Reaction Schemes. IJCSNS International Journal of Computer Science and Network Security, 7(10, October 2007).

[9]. Adelstein, F., Stillerman, M., & Kozen, D. (2002). Malicious Code Detection For Open Firmware. Paper

presented at the 18th Annual Computer Security Applications Conference (ACSAC '02), IEEE

[10]. Poolsapassit, N., & Ray, I. (2007). Investigating Computer Attacks using Attack Trees. Advances in Digital Forensics III, 242, 331-343.

[11]. Yusof, R., Selamat, S. R., & Sahib, S. (2008). Intrusion Alert Correlation Technique Analysis for Heterogeneous Log. IJCSNS International Journal of Computer Science and Network Security, 8(9)

[12]. Kao, D.-Y., Wang, S.-J., Huang, F. F.-Y., Bhatia, S., & Gupta, S. (2008). Dataset Analysis of Proxy Logs Detecting to Curb Propagations in Network Attacks. Paper presented at the ISI 2008 Workshops.

[13]. Lincoln Lab, M. (1999). 1999 DARPA Intrusion Detection Evaluation Plan [Electronic Version].

[14]. McAfee. (2003). Virus Profile: W32/Lovsan.worm.a [Electronic Version]. Retrieved 23/7/09 from http://home.mcafee.com/VirusInfo/VirusProfile.aspx?key=100547.

[15]. Microsoft. (2003). Virus alert about the Blaster worm and its variants [Electronic Version]. Retrieved 23/7/09 from http://support.microsoft.com/kb/826955.

[16]. Symantec. (2003). W32.Blaster.Worm [Electronic Version]. Retrieved 23/7/09 from http://www.symantec.com/security_response/writeup.jsp?docid=2003-081113-0229-99